# The quantum vacuum: an interpretation of quantum mechanics

**Eduardo V. Flores**
Department of Physics & Astronomy
Rowan University
Glassboro, NJ 08028
Email: flores@rowan.edu

**Abstract.** We assume that particles are point-like objects even when not observed. We report on the consequences of our assumption within the realm of quantum theory. An important consequence is the necessity of vacuum fields to account for particle dynamics. The zero-point energy is evidence of the reality of vacuum fields. The dynamics of vacuum fields depends on wave equations, boundary conditions and external sources. The wavefunction is an expression of vacuum fields. Particle events are described by probabilistic distribution functions obtained from vacuum fields. We solve the measurement problem by noticing that in our model the wavefunction does not collapse.

**Introduction**

Our assumption that particles are always point-like objects with actual trajectories has already been considered in Bohmian mechanics [1]. Bohmian mechanics is considered by some to be a paradox-free alternative to the standard theory of quantum mechanics. The standard theory of quantum mechanics consists on the well-known mathematical model interpreted according to the Copenhagen interpretation [2]. However, Bohmian mechanics success is restricted to the non-relativistic level. Bohmian mechanics has an additional mathematical ingredient known as the guiding equation. The guiding equation describes particle trajectory. A difference between our work and Bohmian mechanics is that we do not add extra ingredients to the mathematical structure of quantum mechanics.

On the mathematical aspect of quantum physics much progress has been made at both fronts non-relativistic and relativistic levels. New computational tools allow better predictions for non-relativistic effects [3]. At the relativistic level the Standard model of particles is successful [4]. At the even higher energy, string theories represent a remarkable development [5]. However, development of the interpretation quantum mechanics has not been as significant as its mathematical development. The proof is our inability to find a satisfactory solution to quantum paradoxes such as wave-particle duality, Schrödinger's cat, etc [2]. Some paradoxes are related to the measurement problem in quantum mechanics. Solutions to the measurement problem proposed by "many worlds interpretation", "consistent histories", and Bohmian mechanics are not widely accepted [6].

The Copenhagen interpretation of quantum mechanics emphasizes the role of the observer in a physical outcome: "No elementary phenomenon is a real phenomenon until it is an observed phenomenon [7]." Particles do not have definite physical properties until they are observed. Particles which are not observed do not have trajectories. Wave phenomenon has been observed for matter and light; however, the wavefunction that describes it is considered no more than a mathematical tool for calculations. Attempts to give physical meaning to the wavefunction run quickly into trouble. For instance, at measurement the wavefunction is supposed to collapse instantaneously, thus, it does not obey Schrödinger's equation [2]. We would expect that a wavefunction with physical interpretation would at least obey Schrödinger's equation in all circumstances.

We propose that a consistent interpretation of quantum mechanics requires that particles are truly point-like objects at all times. In this paper we list the consequences of our proposal using well known examples. Our examples are at the non-relativistic level; however, the theory is applicable to the relativistic level too. Our purpose is to present our assumption that particles are point-like even when not observed and its consequences in the clearest possible way. We touch on critical aspects of quantum mechanics such as the uncertainty principle, complementarity, Bell's inequalities and measurement.

**Particle in a box**

We use examples to uncover consequences of our assumption, that particles are point-like at all times. We start with a particle of mass $m$ in a one-dimensional box of size $a$ that has impenetrable walls [2]. The potential is infinite outside and zero inside the box. Inside the box the particle is free. The time independent components of the eigenfunctions, defined inside the box, are

$$\phi_j = \sqrt{\frac{2}{a}} \sin \frac{\pi j x}{a} \qquad j = 1, 2, 3, \ldots \tag{1}$$

The corresponding energy levels are

$$E_j = \frac{\pi^2 \hbar^2 j^2}{2ma^2}. \tag{2}$$

For each energy state $j$ the probability density function, $P_j = \phi_j^* \phi_j$, is

$$P_j = \frac{2}{a} \left( \sin \frac{\pi j x}{a} \right)^2. \tag{3}$$

Let us consider the state that corresponds to $j = 2$. The eigenfunction

$$\phi_2 = \sqrt{\frac{2}{a}} \sin \frac{2\pi x}{a} \tag{4}$$

has a wavelength equal to the length of the box; thus, there are three nodes, one at each wall and one at the center. We plot the corresponding probability density function $P_2$ in Fig. 1. We notice that the particle will not be found at the center of the box. We would like to understand how a free point-like particle is excluded from the center of the box.

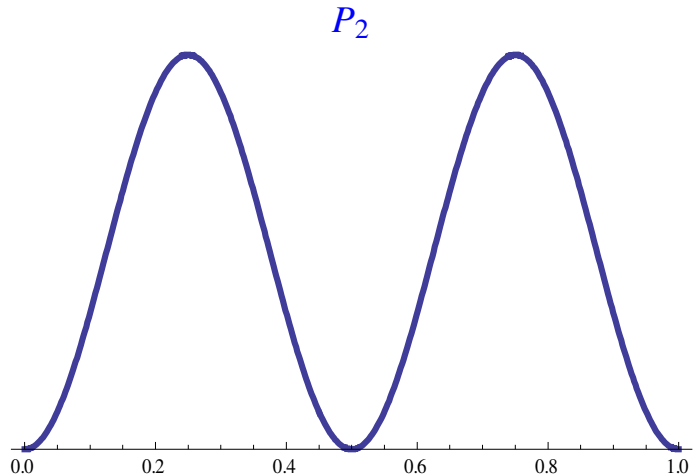

**Fig. 1** Probability density function $P_2$ for a particle in a box with impenetrable walls in state $\phi_2$. Inside the box the point-like particle is free; however, we notice that at the center of the box the particle will not be found as if there was a repulsive force on the particle.

To explain why the particle in state $j = 2$ cannot be found at the center of the box we invoke particle-wave duality [8]. When the particle is not being observed, it does not have to be point-like. In fact, it could be identified with the wave $\phi_2$ in Eq. (4). Since the wavefunction $\phi_2$ is zero at the center of the box and $P_2$ is built from $\phi_2$ then the particle will not be found at the center of the box in an actual measurement. However, we cannot invoke particle-wave duality in our work since we assume that the particle is point-like even when it is not observed. Therefore, the wavefunction $\phi_2$ is by no means the particle. Thus, the first consequence of our basic assumption is that *particle and wave are two separate realities*.

The shape of the probability density function $P_2$ in Fig. 1 could be verified experimentally by repeated measurements of particle position under identical conditions. If we assume that the particle is always point-like we must conclude from the plot in Fig. 1 that inside the box the particle is under an effective force field. For instance, if we find that the particle is never found at the center of the box we must attribute it to a repulsive force in that area. Since the probability density function $P_2$ is built out of the wavefunction $\phi_2$, in Eq. (4), we can conclude that the wavefunction is a manifestation of that force field that acts on the particle.

We would like to find out what generates the wavefunction $\phi_2$ in Eq. (4). In linear theories, like quantum mechanics or electrodynamics, particle dynamics depends on external fields or external forces at least at first order [9]. Since the wavefunction $\phi_2$ represents a force field on the point-like particle it must be an external field. Thus, we deduce that our *particle does not generate $\phi_2$*. We must look for another source for $\phi_2$. We notice that inside the box the particle is free, thus, it is not in contact with any object or external force except for the vacuum. Therefore, the vacuum cannot be truly empty; there must be fields that are intrinsic to the vacuum. We call them vacuum fields. The wavefunction $\phi_2$ must be an expression of vacuum fields. Thus, we observe that *the sources of the wavefunction $\phi_2$ are vacuum fields*.

Vacuum fields are intrinsic components of the vacuum, thus, they must be present even in the absence of particles. Since eigenfunctions $\phi_j$, in Eq. (1), are an expression of vacuum fields, then eigenfunctions $\phi_j$ are present even in the absence of particles. We notice that out of all the states $\phi_j$ the particle would occupy the one that corresponds to its energy $E_j$, in Eq. (2); other states would remain empty. We see that in our example the wavefunction is always present; thus, *there is no collapse of the wavefunction*.

The particle in the box example shows us that vacuum fields, expressed as eigenfunctions $\phi_j$, in Eq. (1), are determined by boundary conditions of the box. The finite size of the box results in quantized states with quantum number $j$. For a given state $j$ the size of the box, $a$, affects the energy of the state, $E_j$ in Eq. (2), and the corresponding wavelength of eigenfunction $\phi_j$. We emphasize that the particle does not determine or generate its own wavefunction; thus, *the actual particle location should not affect the wavefunction*.

Another consequence of the assumption that particles are point-like at all times is *the impossibility that a particle is in a superposition of states*. A point-like particle can only have a single momentum at a given time. At a given time the particle in the box must be moving to the left or to the right. If at a certain time a point would be found left and right of the original point it would cease to be a point since it would become two points. However, the wavefunction can be in a superposition of states. For the particle in a box, the eigenfunction $\phi_2$, in Eq. (4), is a standing wave or a superposition of two identical waves one moving to the left and the other to the right. Once again we see that point-like particle and wave are two separate physical realities.

If we believe that a series of identically prepared experiments with identical measurement conditions will reproduce the plot in Fig. 1 confirming that the particle is in state $\phi_2$, in Eq. (4), then, we must conclude that the location of the particle within the box is random with a probability distribution. This helps us to realize that there is built-in randomness in the physical world. Therefore, vacuum fields alone cannot fully determine the location of the particle; however, we can obtain probabilistic information from vacuum fields.

In quantum mechanics, Heisenberg uncertainty principle states by precise inequalities that certain pairs of physical properties, such as position and momentum, cannot be simultaneously measured with arbitrarily high precision. That is, the more precisely one property is measured, the less precisely the other can be measured. The uncertainty inequality for position and momentum is

$$\Delta x \Delta p \geq \hbar/2. \qquad (5)$$

Thus, for the particle in a box the uncertainty in position, $\Delta x = a$, constrains the uncertainty in momentum, $\Delta p \geq \hbar/(2a)$. If we want to verify that the particle is point-like we would let the size of box go to zero, $a \to 0$, then the uncertainty in momentum would go to infinity, $\Delta p \to \infty$. However, in our theory we assume that the particle is point-like at all times without a need to verify it by direct measurement; thus, no infinite uncertainties are introduced in our theory. The uncertainty principle does not deal with what a particle actually is but with the accuracy when measuring its conjugate properties. Therefore, *our theory of point-like particles at all time is compatible with Heisenberg's uncertainty principle*.

We could generalize our particle in a box example to an example with a more general Hamiltonian, $H$. The potential and boundary conditions would determine the complete set of eigenfunctions $\psi_j$. These eigenfunctions would represent vacuum fields that a particle would experience as force fields. If time dependent potentials, sources, or boundary conditions were present then the overall argument would not change. The main difference from the time independent case studied above would be that the vacuum fields would change in time and in general would propagate according to Schrödinger's equation in the non-relativistic case. *Vacuum fields are dynamic even in the absence of a particle to occupy them*.

**Zero-point energy**

*"The vacuum holds the key to a full understanding of the forces of nature"* [10].

We saw before that if the particle is point-like at all times then the wavefunction has physical reality. Furthermore, the wavefunction is an expression of intrinsic vacuum fields. To learn more about vacuum fields we analyze them from another perspective.

Our work applies to relativistic and non-relativistic quantum mechanics. However, we explain our model using a non-relativistic many particle system. In a non-relativistic many particle system we have at least two *equivalent* methods to deal with a problem: the many-body approach and the second quantization method. In the many-body approach we build the wavefunction of the system as a product of 1-particle wavefunctions and carefully handle the symmetry properties of the particles. In the second quantization approach we introduce creation and annihilation operators that by their algebraic rules automatically do the bookkeeping necessary to preserve the symmetry of the wavefunction [11]. The second quantization method also provides evidence of intrinsic vacuum fields.

There is excellent literature on the second quantization procedure [11,12]. We only mention key aspects to be used in our argument. We consider the non-relativistic case described by Schrodinger's equation. Let $\psi_j$ be the complete set of eigenfunctions of the Hamiltonian $H$. The eigenfunctions $\psi_j$ are determined by the potential and boundary conditions. The field operator is defined as

$$\Psi = \sum_j c_j \psi_j \qquad (6)$$

where the $c_j$ are destruction operators and $\psi_j$ are field modes or eigenfunctions of the Hamiltonian. Let $|n\rangle_j$ be the state of $n$ particles in mode $\psi_j$. If we operate on a one-particle state with a destruction operator we obtain the vacuum $c_j|1\rangle_j = |0\rangle$. Operating on the vacuum with a creation operator $c_j^\dagger|0\rangle = |1\rangle_j$ results in a particle in mode $j$. To obtain the eigenfunction $\psi_j$ for this particle we use the field operator $\Psi$ in Eq. (6),

$$\langle 0|\Psi|1\rangle_j = \psi_j. \tag{7}$$

"The second quantization approach exposes the fact that there is no vacuum in the ordinary sense of tranquil nothingness" [12]. There is instead zero-point energy for every field mode $\psi_j$. We see this by writing a symmetric Hamiltonian in terms of field operators and the result is

$$H = \sum_j E_j(c_j^\dagger c_j + 1/2), \tag{8}$$

where $E_j$ is the eigenvalue of eigenfunction $\psi_j$. The second term in Eq. (8) shows that even in the absence of particles there is energy $\frac{1}{2}E_j$ associate with every field mode $\psi_j$. The energy in the vacuum is known as the zero-point energy. Casimir's effect, Lamb's shift and spontaneous emission are confirmations of the reality of zero-point energy, $\frac{1}{2}E_j$, and corresponding field mode $\psi_j$ [12].

Casimir's effect is a net attractive force present between parallel plates in vacuum. Contributions to Casimir's effect come from every mode $\psi_j$ of every quantum field [12]. Let us consider just the case of Casimir's effect due to electromagnetic field. Imagine two parallel plates with infinite conductivity so that no electromagnetic field penetrates the plates. Boundary conditions require that the electromagnetic field at the plates is zero. This implies that modes that do not meet this boundary condition are not allowed within the plates. On the other hand, outside the plates these modes are not excluded. Thus, the difference in modes on each side of a plate results in a net pressure. This leads to an attractive force between plates.

From our analysis of Casimir's effect due to electromagnetic field we see that field modes are determined by boundary conditions of the problem. For instance, modes that do not meet boundary conditions are not present in the vacuum within the plates. Casimir's effect provides evidence that these modes $\psi_j$ are real; otherwise, there would be no attractive force between plates. Our example demonstrates that *the vacuum can be manipulated*. We emphasize that field modes, intrinsic to the vacuum, are determined by potentials and boundary conditions.

Therefore, after all matter and radiation have been removed from a region of spacetime what remains is a physical vacuum which includes field modes $\psi_j$ with corresponding zero-point energy $\frac{1}{2}E_j$. We propose that field modes, $\psi_j$, play a key role in the dynamics of particles. In other words, field modes or a linear combination of them are responsible for wave effects such as the one in Fig. 1.

Returning to the particle in a box example we recall that eigenfunction $\phi_2$, in Eq. (4), is not the particle and it is not generated by the particle, but it is an expression of vacuum fields. Now, we go further and *identify eigenfunction $\phi_2$ of the particle in a box example with the $j = 2$ mode of the vacuum,*

$$\phi_2 \equiv \psi_2. \tag{9}$$

Similarly, we identify eigenfunction $\phi_j$ from the particle in the box example with the $j$ mode of the vacuum, $\phi_j \equiv \psi_j$. Therefore, the vacuum fields we require in the particle in a box example are actually the field modes of the vacuum in the second quantization technique.

We keep in mind that the second quantization technique is not essential as there are other techniques to deal with many body problems at relativistic and non-relativistic levels [11]. An advantage of the second quantization technique is that by treating a field mode as a harmonic oscillator it reveals the existence of zero-point energy which shows that its corresponding mode exists even in the absence of particles. Regardless of how we obtain field modes, we identify fields modes described by eigenfunctions, $\psi_j$, as the vacuum fields that drive particle dynamics.

**Particle and the two slits experiment**

We apply our observations above to our next example: the particle and two slits experiment that according to Richard Feynman contains the *only* mystery of quantum mechanics [13]. Thus, let us consider a very large absorbent wall with two pinholes. The potential everywhere else is zero. Let us assume that the source of the particles is at infinity so that the source does not affect our system.

To determine the situation for vacuum fields before a particle arrives we consider boundary conditions and the free Schrödinger's equation for the non-relativistic case. Right and left of the wall there will be field modes represented by plane waves with minor disturbances in the region near the pinholes. However, there will be a mode moving to the right with the precise wavelength that will produce a picture like the one in Fig. 2. Left of the wall it will be a plane wave. Right of the wall it will be two cones emanating out of the pinholes. The cones will intersect and form regions of constructive and destructive interference. Of course, if one of the pinholes were closed, then the boundary conditions would have changed and we would only observe a single cone and no interference.

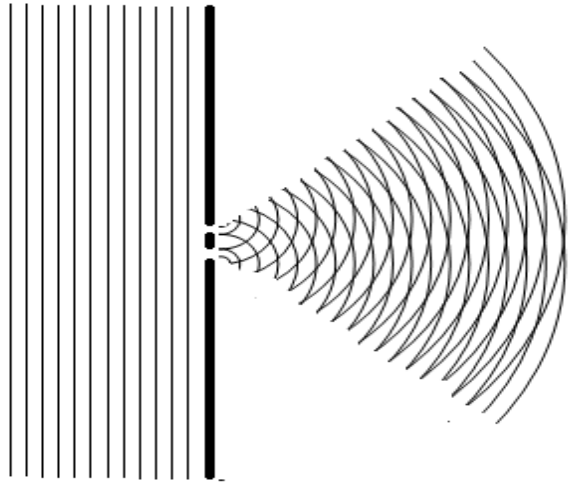

**Fig. 2** A particular field mode shaped by boundary conditions, an absorbent wall with two pinholes. Left of the wall we have a plane wave. Right of the wall there is a region of constructive and destructive interference. This mode could be occupied by a particle coming from infinity left of the wall.

Suppose that a particle that comes from far away approaches the wall from the left. If the particle momentum is right it will occupy the mode depicted in Fig. 2. If the particle goes through one of the two pinholes it would eventually hit a screen located far away right of the wall. Along its path, the particle would be driven by the field mode; thus, it would be kept away from regions of destructive interference and it would be lead to regions of constructive interference. Thus, it is more likely that the particle would hit the screen at a place where constructive interference occurs. The hit would be recorded as a point on the screen.

When the particle hits the screen it leaves a mark which constitutes a measurement. At measurement, the particle vacates the field mode described in Fig. 2 and occupies a field mode generated by elements that form the screen. The particle could end up bounded to an atom on the screen. Information is generated when the particle changes field mode and there is a corresponding energy transfer. In many cases energy transfer is reflected by the emission or absorption of photons. We note that the mode in Fig. 2 will be unchanged even after being vacated by the particle. If another particle with identical momentum to the first approaches the wall it will occupy the mode described in Fig. 2. The particle could be detected as another point on the screen. The collection of many identically prepared particles, even if they come one at a time, would eventually form on the screen an interference pattern with bright and dark fringes.

If we decide to investigate through which pinhole a particle goes through we could block one of the pinholes. When a pinhole is blocked we have changed boundary conditions which results in a different set of field modes. Now, right of the wall, there will be a single cone that by itself produces no interference. However, we may decide to obtain information about which pinhole the particle crosses using a mechanism that partially blocks the pinholes. Imagine that the particles that come from far away are electrons and a source of photons is placed behind the pinholes. The setup is such that whenever a scattered photon is detected we know which pinhole the electron crossed [13]. As it is well known this does not work either. The uncertainty principle keeps us from getting information about which pinhole each electron crosses through while simultaneously maintaining an interference pattern.

Richard Feynman questions the existence of a mechanism that would explain the odd results of the particle and the two slits experiment [13]. According to classical physics the probability density obtained with two slits open is the sum of probability densities obtained by opening one slit at a time. In quantum mechanics it is different; the probability density obtained with two slits open is not the sum of probability densities obtained by opening one slit at a time. This is Feynman's *only* mystery of quantum mechanics. It appears to us that this mystery of quantum mechanics has been solved. Our answer is found in the vacuum. Intrinsic vacuum fields are responsible for wave effects seen in this experiment. In our case we have two different boundary conditions: one pinhole open and two pinholes open. Two different boundary conditions generate two different vacuum field configurations. Particles are directly affected by vacuum fields; thus, we get two different results for two different setups.

**Immediate consequences**

*Complementarity*
We may wonder how Bohr's principle of complementarity and our theory could possibly agree. In our theory we already know that the particle is point-like, thus, every time we observe wave phenomenon we seem to have both particle and wave. This is not surprising. In any two-slit type experiment performed with one particle at a time we end up seeing both: dots on a screen (particle) and the formation of an interference pattern (wave). However, complementarity is tied to direct observation of particle and wave effects. In quantum mechanics direct observation always affects the object being observed. Direct observation of particle and wave effects are constrained by either the uncertainty principle [14] or quantum entanglements [15]. In the two-slit type experiment, we could attempt to directly measure through which pinhole a particle goes through and simultaneously measure interference effects. The results of this experiment would be constrained by the complementarity inequality,

$$K^2 + V^2 \leq 1, \qquad (10)$$

where $K$ is the which-way information (particle) parameter and $V$ is the visibility (wave) parameter [16]. Such is the situation for standard quantum mechanics and also for our model. Therefore, within our model it is not possible to *directly* observe both particle and wave aspects beyond the complementarity inequality, Eq. (10).

*Bell's inequalities*
We mentioned before that in our theory the particle is not the source of the wavefunction. This implies that the particle unique or definite position, momentum, spin state, etc. will not affect the development of the wavefunction. The wavefunction develops due to changes in boundary conditions, external sources, and potentials. We also mentioned before that in our theory the wavefunction acts on the particle as if it was an external field; thus, the wave function is the driving force. Therefore, any result obtained from our theory concerning measurements suggested by Bell's inequalities would line up with the quantum mechanical expectation which is based on the wavefunction [17,18]. Our theory and the mathematical formalism of quantum mechanics are actually the same. Thus, *our theory represents a form of realism that agrees with quantum mechanics in a Bell's inequality type experiment* [2].

*Tracing the particle path*
If the particle is a point-like object at all times we could in principle follow its trajectory. One approach to trace the path of the particle uses energy-momentum conservation. This technique has been used before in the analysis of Wheeler's delayed choice experiment [19]. A laser beam enters a Mach-Zehnder setup as in Fig. 3. The beam splits and takes two different paths which at some point cross. At the end of each path there is a detector. When a single photon is present only one detector clicks. Applying energy-momentum conservation we extrapolate the path of the photon all the way to the initial beam. However, this technique is not in agreement with standard quantum mechanics. The reason is that in standard quantum mechanics the path of a particle before detection is meaningless [20]. In fact, the particle before detection consists of two beams: the one in front of the detector that clicks and the one in front of the other detector that does not click. The click of detector 1 produces an abrupt collapse of the wavefunction to a point in detector 1. Even though the particle was detected at detector 1 we cannot determine where it was even an instant before detection. Thus, the click of detector 1 gives no path information at all according to the standard interpretation of quantum mechanics.

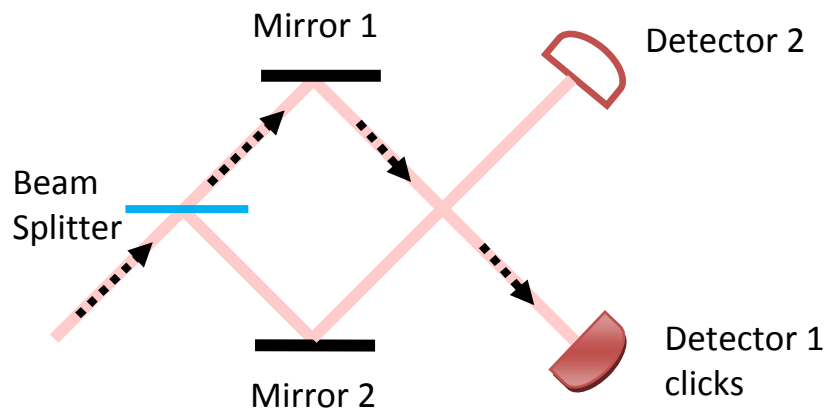

**Fig. 3** Mach-Zehnder set up for the delayed choice experiment. If detector 1 clicks we may trace the path of the photon applying energy-momentum conservation. However, this technique assumes that the particle is point-like before being observed by the click of the detector.

In spite of conflicts with the standard interpretation of quantum mechanics the conservation laws technique is heavily used in experimental high energy physics. In high energy experimental physics, important results have been obtained by extrapolating the path of a particle using the conservation laws technique [21]. However, applying the conservation laws technique to obtain the path of the particle before detection is a sound technique within our interpretation. In our interpretation the particle is always point-like; thus, when detector 1 in Fig. 3 clicks then by energy-momentum conservation we trace unambiguously the particle path all the way to where it enters the setup. We have a definite measurement. The success of the conservation laws technique, in high energy physics, supports our interpretation of quantum mechanics over the standard interpretation.

*Resolution of the measurement problem*
According to our theory, most measurements amount to particles changing state; particles vacate one state and occupy another. This situation resembles the case of a bound electron making transitions from state to state. The change of state of a particle is marked by energy transfer that could be used to obtain information. Therefore, we view measurement as a regular quantum process.

We saw before that if particles are point-like at all times then they are always in a unique state. Particle state can change but it is always a single state. A measurement shows the state of the particle. On the other hand, the wavefunction is normally in a superposition of states. However, measuring the state of a particle does not necessarily change the wavefunction; the wavefunction would change only if there were changes in potentials, external sources or boundary conditions. Thus, at measurement, the wavefunction does not collapse when a particle is detected, absorbed or trapped. The collapse of the wavefunction is bound to cause problems since it does not obey Schrödinger's equation. However, in our model, changes in the wavefunction *always* obey Schrödinger's equation; thus, the measurement problem does not emerge.

## Conclusions

If particles are truly point-like objects even when they are not observed, then the wave aspect of quantum mechanics acquires physical reality independent of the particle. The consistency of our interpretation is a strong argument in favor of particles being point-like objects all the time. We believe that to make progress in physics *it is important to know what particles actually are*. Knowledge about what particles are will eventually help us to make progress in our understanding of the physical world.

Our work shows that *quantum mechanics is basically a theory that studies the development of vacuum fields*. The importance of vacuum fields is that they are the fields where particles dwell and move. Moreover, these fields exert forces on particles. The wavefunction $\psi$ is a manifestation of that force field that acts on a particle. The probability density function, $P = \psi^*\psi$, reflects that force field since it is built out of $\psi$.

Our model is mathematically as consistent as the mathematical consistency of standard quantum theory. What our model represents is nothing more than an interpretation of particles and wavefunctions. It appears to us that our interpretation is better than the standard interpretation as it resolves known paradoxes and opens up new directions of research.

**References**

[1] D. Durr, S. Goldstein, R Tumulka, and N Zangh, "Bohmian Mechanics," Compendium of Quantum Physics: Concepts, Experiments, History and Philosophy, Edited by D Greenberger, K. Hentschel, and F. Weinert, (Springer, 2009)

[2] A. Goswami, Quantum Mechanics (Wm. C. Brown Publishers, Dubuque IA, 1992)

[3] L. Marchildon, Quantum Mechanics: From Basic Principles to Numerical Methods and Applications (Springer, New York, 2010)

[4] R. Mann, An Introduction to the Standard Model of Particle Physics (CRC Press, 2009)

[5] J. Polchinski, String Theory, Vol. 1 (Cambridge University Press, 2005)

[6] D. Wallace, The Quantum Measurement Problem: State of Play, chapter 1 of D. Rickles (ed), The Ashgate Companion to the New Philosophy of Physics (Ashgate, 2008) **arXiv:0712.0149**

[7] J. A. Wheeler (2), "The 'Past' and the 'Delayed-Choice Double-Slit Experiment'," pp 9–48, in A.R. Marlow, editor, Mathematical Foundations of Quantum Theory (Academic Press, 1978)

[8] R. Eisberg, R. Resnick, Quantum Physics of Atoms, Molecules, Solids, Nuclei, and Particles (Wiley, New York, 1974)

[9] L. D. Landau and E. M. Lifshitz, The Theory of Classical Fields, Vol. 2, (Pergamon Press, 1975)

[10] P. C. W. Davies, Superforce (Simon and Schuster, New York, 1985) p. 104

[11] J. D. Bjorken [and] Sidney D. Drell, Relativistic quantum fields, (McGraw-Hill, New York, 1965)

[12] P. W. Milonni, Quantum vacuum : an introduction to quantum electrodynamics, (Academic Press, INC, San Diego, CA, 1994)

[13] R.P. Feynman, The Feynman Lectures on Physics, Vol. 3, (Addison-Wesley, Reading MA, 1963), pp. 1-9.

[14] S. Dürr, and G. Rempe, "Can wave–particle duality be based on the uncertainty relation?" Am. J. Phys. 68, 1021 (2000).

[15] B.-G. Englert, "Fringe visibility and which-way information: An inequality," Phys. Rev. Lett. 77 2154 (1996)

[16] D. Greenberger and A. Yasin, "Simultaneous wave and particle knowledge in a neutron interferometer," *Phys. Lett. A* **128**, 391 (1988).

[17] J. S. Bell, Physics 1, 195 (1964)

[18] J. F. Clauser, and M. A. Horne, Phys. Rev. 10, 526 (1974)

[19] J. A. Wheeler, "Law Without Law," Quantum Theory and Measurement (Princeton: Princeton University Press, 1983).

[20] Flores, E., De Tata, J., Foundation of Physics online (June, 2010)   DOI: 10.1007/s10701-010-9477-4
   [arXiv:1001.4785](arXiv:1001.4785)

[21] R. Fernow, Introduction to Experimental Particle Physics, (Cambridge University Press, 1989)